\documentstyle[amssymb,aps,12pt]{revtex}

\begin{document}
\title{Quantum Phase Transition of Condensed Bosons in Optical Lattices}
\author{Jun-Jun Liang$^{1,}\thanks{%
Corresponding author$^{\prime }$s E-mail address: liangjj929@yahoo.com.cn}$%
,J.-Q. Liang$^1$and W. -M. Liu$^2$}
\address{$^1$Institute of Theoretical Physics, Shanxi University, Taiyuan 030006,\\
China\\
$^2$Institute of Physics and Center for Condensed Matter Physics, Chinese\\
Academy of Sciences, P. O. Box 2101,\ \\
Beijing 100088, China}
\maketitle

\begin{abstract}
In this paper we study the superfluid-Mott-insulator phase transition of
ultracold dilute gas of bosonic atoms in an optical lattice by means of
Green function method and Bogliubov transformation as well. The
superfluid-Mott insulator phase transition condition is determined by the
energy-band structure with an obvious interpretation of the transition
mechanism. Moreover the superfluid phase is explained explicitly from the
energy spectrum derived in terms of Bogliubov approach.
\end{abstract}

PACS number(s): 03.75.Lm, 67.40.-w, 39.25.+k, 32.80.Pj 

\section{INTRODUCTION}

Ultracold bosons in an optical lattice provide a tunable quantum system with
variance of the potential depth and lattice constant which can be achieved
by adjusting the parameters of the configuration of laser beams. Various
quantum phenomena, for instance, Bloch oscillations, Wannier-Stack ladders%
\cite{O. Morsch} have been investigated in such a system which shares
spatial periodicity with crystal lattice in solid state physics, however is
immune from scattering of impurities or phonons. The
superfluid-Mott-insulator (SMI)\ phase transition is one of the most
significant quantum phenomena of condensate bosons in the optical lattice. A
known analogous system exhibiting the SMI phase transition is liquid helium
with short-range repulsive interaction in periodic potential\cite{Matthew P.
A. Fisher}. The atomic gas in Bose-Einstein condensate (BEC) subjected to
the lattice potential which is turned on smoothly can be kept in the
superfluid phase (SFP) as long as the repulsive interaction between atoms is
small with respect to the tunnel coupling. With increasing of the potential
depth of the optical lattice it is getting more and more difficult for
bosons to tunnel from one site to the other, and finally the system attends
an insulator phase above a critical value of the potential depth.
Considerable attention has been attracted to theoretical researches for
understanding of the phase transition and determining of the transition
condition\cite{D. Jaksch}-\cite{D. van Oosten}, in which the Bose-Hubbard
Model(BHM) is introduced as the starting point of the theoretical studies%
\cite{D. Jaksch}. The phase transition phenomena have been also observed
experimentally in BEC loaded in a three dimensional optical lattice\cite{D.
Jaksch}. Using a strong coupling expansion in terms of the hopping term
called the decoupling approximation, which is as a matter of fact based on
the mean-field method D. van Oosten et. al have obtained an analytic phase
transition condition (see the following Eq.(\ref{condition})). From an
alternative viewpoint the phase transition condition can be also determined
from the energy spectrum of the system since the excitation spectrum is
neccesarily gapless for the SFP while has a finite gap for the Mott
insulator phase (MIP). We in the present paper use both the Green function
method and Bogliubov approach to obtain the explicit excitation energy
spectrum and hence to investigate the SMI phase transition.

We begin with the following second-quantized Hamiltonian operator\cite{D.
Jaksch} for the system of bosonic atoms in the optical lattice, 
\begin{eqnarray}
\widehat{H} &=&\int d{\bf x}\widehat{\psi }^{\dagger }\left( {\bf x}\right)
\left( -\frac{\hslash ^2}{2m}\triangledown ^2+V_0\left( {\bf x}\right) +V_T(%
{\bf x})\right) \widehat{\psi }\left( {\bf x}\right)  \nonumber \\
&&+\frac g2\int d{\bf x}\widehat{\psi }^{\dagger }\left( {\bf x}\right) 
\widehat{\psi }^{\dagger }\left( {\bf x}\right) \widehat{\psi }\left( {\bf x}%
\right) \widehat{\psi }\left( {\bf x}\right)
\end{eqnarray}
where $\widehat{\psi }\left( {\bf x}\right) $ and $\widehat{\psi }^{\dagger
}\left( {\bf x}\right) $ denote the boson field operators which obey the
boson commutation relation 
\begin{equation}
\left[ \widehat{\psi }\left( {\bf x}\right) \text{, }\widehat{\psi }%
^{\dagger }\left( {\bf x}^{\prime }\right) \right] =\delta \left( {\bf x-x}%
^{\prime }\right) \text{.}
\end{equation}
Here $V_0\left( {\bf x}\right) =$ $\sum_{j=1}^3V_{j0}\left( {\bf x}\right)
\sin ^2(\frac{2\pi x_j}\lambda )$ is the potential of the optical lattice
formed by the laser light of wavelength $\lambda $, and hence the lattice
constant is $d=\lambda /2$. $V_T({\bf x})$ denotes an external trap
potential, and the interparticle interaction is approximated by the
short-range potential $g\delta \left( {\bf x-x}^{\prime }\right) $, where $%
g=4\pi a_s\hslash ^2/m$ is the coupling constant with $a_s$ the s-wave
scattering length. Expanding the field operator $\widehat{\psi }\left( {\bf x%
}\right) $ in the Wannier basis such that $\widehat{\psi }\left( {\bf x}%
\right) =\sum_i\widehat{a}_iw\left( {\bf x-x}_i\right) $, we obtain the
Bose-Hubbard model 
\begin{equation}
\widehat{H}=-J\sum\limits_{\left\langle i,j\right\rangle }\widehat{a}%
_i^{\dagger }\widehat{a}_j+\sum\limits_i\varepsilon _i\widehat{n}_i+\frac 12%
U\sum_i\widehat{n}_i(\widehat{n}_i-1)  \label{Hamiltonian}
\end{equation}
where $\widehat{a}_i$ is the annihilation operator of a particle at the
lattice site $i$, which is assumed as being in a state described by the
Wannier function $w({\bf x}-{\bf x}_i)$ of the lowest energy band localized
on $ith$ site. This implies the assumption that the energy involved in the
system is small comparing with the excitation energies of the second band. $%
{\bf x}_i$ denotes the position of the $ith$ local minimum of the optical
potential, and $\widehat{n}_i=\widehat{a}_i^{\dagger }\widehat{a}_i$ is the
number operator. The annihilation and creation operators $\widehat{a}_i$ and 
$\widehat{a}_i^{\dagger }$ obey the canonical commutation relations $[%
\widehat{a}_i$ $,\widehat{a}_i^{\dagger }$ $]$ $=$ $\delta _{ij}$. The
parameter $J$ is the hopping matrix element between adjacent sites $i$, $j$,
and is evaluated as

\begin{equation}
J=-\int d{\bf x}w^{*}({\bf x}-{\bf x}_i)[-\frac{\hbar ^2}{2m}\nabla ^2+V_0(%
{\bf x})]w({\bf x}-{\bf x}_j)\text{.}  \label{J}
\end{equation}
The energy offset of each lattice site, $\varepsilon _i=\int d{\bf x}V_T(%
{\bf x})\left| w({\bf x}-{\bf x}_i)\right| ^2\thickapprox V_T({\bf x}_i)$ ,
is assumed to be of the same value $\varepsilon $ in the present paper. The
interparticle interaction is characterized by the parameter

\begin{equation}
U=g\int d{\bf x}|w({\bf x})|^4  \label{U}
\end{equation}

For the sake of convenience we rewrite the Hamiltonian(\ref{Hamiltonian}) in
the following form:

\begin{equation}
\widehat{H}=\widehat{H}_0+\frac 12U\sum_i\widehat{n}_i(\widehat{n}%
_i-1),\quad \widehat{H}_0=\sum\limits_{i,j}T_{ij}\widehat{a}_i^{\dagger }%
\widehat{a}_j  \label{Hamiltonian1}
\end{equation}
where

\[
T_{ij}=\left\{ 
\begin{array}{c}
\varepsilon \qquad \qquad \qquad \text{for }i=j \\ 
-J\qquad i,j\text{ are nearest neighbors} \\ 
0\qquad \qquad \qquad \qquad \text{otherwise}
\end{array}
\right. 
\]
We see that the first part in Hamiltonian Eq.(\ref{Hamiltonian1}), $\widehat{%
H}_0$, is the same as that of a simple lattice under tight-binding
approximation (TBA) in solid state physics, so we can rewrite $T_{ij}$ as

\begin{equation}
T_{ij}=N_s^{-1}\sum_k\varepsilon (k)\exp [i{\bf k}\cdot ({\bf x}_i-{\bf x}%
_j)],
\end{equation}
where ${\bf k}$ is the wave vector in the first Brillouin zone, $N_s$ is the
total number of the lattice sites and $\varepsilon (k)$ is the energy
spectrum of the Hamiltonian $\widehat{H}_0$. The inverse transformation is
written as

\begin{equation}
\varepsilon (k)=N_s^{-1}\sum_{i,j}T_{ij}\exp [-i{\bf k}\cdot ({\bf x}_i-{\bf %
x}_j)]  \label{energy}
\end{equation}
and can be approximated by $\varepsilon (k)\approx \varepsilon -\frac J{N_s}%
\sum\limits_{<ij>}$ $exp[-i{\bf k}\cdot ({\bf x}_i-{\bf x}_j)]$ (TBA energy
band) in simple cubic lattice. The explicit energy spectrum is seen to be

\begin{equation}
\varepsilon (k)=\varepsilon -Jz\cos (\frac{k\lambda }2)
\end{equation}
where z is the number of nearest neighbors of each site.

The existence of a finite gap in the excitation spectrum is the
characteristic of the MIP. In section II we attempt to determine the SMI
phase transition condition from the energy band structure of the ultracold
bosonic atoms in optical lattice in terms of Green function method. In
section III the Bogliubov transformation is used to obtain exact energy
spectrum with which the superfluid phase is explained explicitly.

\section{GREEN FUNCTION APPROACH}

We begin with the operators, $\widehat{a}_i(t)$, $\widehat{a}_j^{\dagger
}(t^{\prime })$ , in Heisenberg picture, i.e. $\widehat{a}_i(t)=e^{i\widehat{%
H}t}\widehat{a}_ie^{-i\widehat{H}t}$ and $\widehat{a}_j^{\dagger }(t^{\prime
})=e^{i\widehat{H}t^{\prime }}\widehat{a}_j^{\dagger }e^{-i\widehat{H}%
t^{\prime }}$ (in the unit of $\hbar =1$). The retarded single-particle
Green function at zero temperature \cite{A. L. Fetter} is defined by 
\begin{eqnarray}
\left\langle \left\langle \widehat{a}_i(t);\text{ }\widehat{a}_j^{\dagger
}(t^{\prime })\right\rangle \right\rangle &=&-i\theta (t-t^{\prime
})\left\langle [\widehat{a}_i(t),\text{ }\widehat{a}_j^{\dagger }(t^{\prime
})]\right\rangle  \label{defination} \\
&=&-i\theta (t-t^{\prime })\left\{ \left\langle \widehat{a}_i(t)\widehat{a}%
_j^{\dagger }(t^{\prime })\right\rangle -\left\langle \widehat{a}_j^{\dagger
}(t^{\prime })\widehat{a}_i(t)\right\rangle \right\}  \nonumber
\end{eqnarray}
where $\theta (t-t^{\prime })$ is the step function:

\[
\theta (t-t^{\prime })=\left\{ 
{1\text{, \qquad }t>t^{\prime }\text{ } \atop 0\text{, \qquad }t<t^{\prime }}%
\right. 
\]
The Green function $\left\langle \left\langle \widehat{a}_i(t);\text{ }%
\widehat{a}_j^{\dagger }(t^{\prime })\right\rangle \right\rangle $ depends
only on the time difference $(t-t^{\prime })$. The Fourier transformation of
the retarded Green function $\left\langle \left\langle \widehat{a}_i(t);%
\text{ }\widehat{a}_j^{\dagger }(0)\right\rangle \right\rangle $ is seen to
be 
\[
G_{ij}(\omega )\equiv \left\langle \left\langle \widehat{a}_i|\widehat{a}%
_j^{\dagger }\right\rangle \right\rangle _\omega =\frac 1{2\pi }%
\int\limits_{-\infty }^\infty dt\left\langle \left\langle \widehat{a}_i(t);%
\text{ }\widehat{a}_j^{\dagger }(0)\right\rangle \right\rangle \exp
[i(\omega +i\eta )]\text{, \qquad }\eta =+0 
\]
for a real frequency $\omega $. Using Heisenberg equation, we obtain

\begin{equation}
\omega G_{ij}(\omega )=\left\langle [\widehat{a}_i\text{, }\widehat{a}%
_j^{\dagger }]\right\rangle +\left\langle \left\langle [\widehat{a}_i\text{, 
}H]|\widehat{a}_j^{\dagger }\right\rangle \right\rangle _\omega
\label{equation1}
\end{equation}
which can be evaluated in terms of the commutation relation,

\begin{equation}
\lbrack \widehat{a}_i\text{, }\widehat{H}]=\sum_jT_{ij}\widehat{a}_j+U%
\widehat{n}_i\widehat{a}_i\quad \text{,}  \label{commutor1}
\end{equation}
where $\left\langle {}\right\rangle $ denotes the ground state expectation
value. The result is

\begin{equation}
\omega G_{ij}(\omega )=\delta _{ij}+\sum_mT_{im}G_{mj}(\omega )+U\Gamma
_{ij}(\omega ),  \label{single green}
\end{equation}
where 
\[
\Gamma _{ij}(\omega )\equiv \left\langle \left\langle \widehat{n}_i\widehat{a%
}_i|\widehat{a}_j^{\dagger }\right\rangle \right\rangle _\omega 
\]
is the higher order Green function which satisfies the following equation

\begin{equation}
\omega \Gamma _{ij}(\omega )=\left\langle [\widehat{n}_i\widehat{a}_i\text{, 
}\widehat{a}_j^{\dagger }]\right\rangle +\left\langle \left\langle [\widehat{%
n}_i\widehat{a}_i\text{, }\widehat{H}]|\widehat{a}_j^{\dagger }\right\rangle
\right\rangle _\omega \quad .  \label{equation2}
\end{equation}
It is easy to find

\begin{equation}
\lbrack \widehat{n}_i\widehat{a}_i,\widehat{H}]=\varepsilon \widehat{n}_i%
\widehat{a}_i+\sum_{j(\neq i)}T_{ij}\underline{\widehat{n}_i}\widehat{a}%
_j+\sum_jT_{ij(}\underline{\widehat{a}_i^{\dagger }\widehat{a}_j-\widehat{a}%
_j^{\dagger }\widehat{a}_i})\widehat{a}_i+U\underline{\widehat{n}_i}\widehat{%
n}_i\widehat{a}_i  \label{commutor2}
\end{equation}
Substituting Eq.(\ref{commutor2}) into Eq.(\ref{equation2}), we see that the
obtained equation is not closed, because more higher order Green functions
appear in the formula of $\Gamma _{ij}(\omega )$. We in this stage use mean
field approximation for the underlined operators in Eq.(\ref{commutor2}),
i.e. the number operator $\widehat{n_i}$ is replaced by its expectation
value $\left\langle \widehat{n_i}\right\rangle $

\begin{equation}
\sum_jT_{ij}\widehat{n}_i\widehat{a}_j\thickapprox \sum_jT_{ij}\left\langle 
\widehat{n}_i\right\rangle \widehat{a}_j\thickapprox n_0\sum_jT_{ij}\widehat{%
a}_j.
\end{equation}
In the above approximation we have assumed that the average occupation
number of Bose atoms condensed on ground state in each site of the optical
lattice is the same, i.e., $\left\langle \widehat{n}_i\right\rangle \equiv
n_0$.

Since the obvious symmetry of $\varepsilon (k)=\varepsilon (-k)$, we have
the equality $T_{ij}=T_{ji}$ according to Eq.(\ref{energy}). Utilizing
translation symmetry of the Bose system we moreover obtain 
\begin{equation}
\sum_jT_{ij}\left( \left\langle \widehat{a}_i^{\dagger }\widehat{a}%
_j\right\rangle -\left\langle \widehat{a}_j^{\dagger }\widehat{a}%
_i\right\rangle \right) =0.  \label{approximation1}
\end{equation}
The Eq. (\ref{commutor2}) is then simplified as

\[
\lbrack \widehat{n}_i\widehat{a}_i,\widehat{H}]\thickapprox (\varepsilon
+Un_0)\widehat{n}_i\widehat{a}_i+n_0\sum_{j(\neq i)}T_{ij}\widehat{a}_j. 
\]
A closed equation for the Green function is seen to be

\begin{equation}
\omega \Gamma _{ij}(\omega )=2n_0\delta _{ij}+n_0\sum_mT_{im}G_{ij}(\omega
)+(\varepsilon +Un_0)\Gamma _{ij}(\omega )  \label{correlation}
\end{equation}
from which we find

\begin{equation}
\Gamma _{ij}(\omega )=\frac{n_0}{\omega -\varepsilon -Un_0}(2\delta
_{ij}+\sum_{m(\neq i)}T_{im}G_{mj}(\omega )).  \label{high order}
\end{equation}
Substitution of $\Gamma _{ij}(\omega )$ in Eq. (\ref{high order}) into (\ref
{single green}) yields 
\begin{equation}
\omega G_{ij}(\omega )=\delta _{ij}+\sum_mT_{im}G_{mj}(\omega )+\frac{Un_0}{%
\omega -\varepsilon -Un_0}(2\delta _{ij}+\sum_{m(\neq i)}T_{im}G_{mj}(\omega
))  \label{site green}
\end{equation}
which is the equation for the site space Green function $G_{ij}(\omega )$.
The Green function $G_{ij}(\omega )$ is a function of the position
difference (${\bf x}_i-{\bf x}_j$) of two sites only for a system with
translational invariance. The equation (\ref{site green}) for the Green
function $G_{ij}(\omega )$ can be solved with the Fourier transformation. To
this end we express the site space operator $\widehat{a}_i$ in terms of the
wave-vector-space operator $\widehat{a}_k$

\begin{eqnarray}
\widehat{a}_i &=&\frac 1{\sqrt{N_s}}\sum_ke^{i{\bf k\cdot x}_i}\widehat{a}_k
\nonumber \\
\widehat{a}_i^{\dagger } &=&\frac 1{\sqrt{N_s}}\sum_ke^{-i{\bf k\cdot x}_i}%
\widehat{a}_k^{\dagger }.  \label{transform}
\end{eqnarray}
We can prove that the single particle Green function in the Bloch
representation is orthogonal, i.e.,

\begin{eqnarray*}
G_{kk^{\prime }}(\omega ) &\equiv &\left\langle \left\langle \widehat{a}_k|%
\widehat{a}_{k^{\prime }}^{\dagger }\right\rangle \right\rangle _\omega \\
&=&\frac 1N\sum_{i,j}e^{-i{\bf k\cdot x}_i}e^{i{\bf k}^{\prime }{\bf \cdot x}%
_j}\left\langle \left\langle \widehat{a}_i|\widehat{a}_j^{\dagger
}\right\rangle \right\rangle _\omega \\
&=&\delta _{_{kk^{\prime }}}G_k(\omega )
\end{eqnarray*}
where $G_k(\omega )=$ $\left\langle \left\langle \widehat{a}_k|\widehat{a}%
_k^{\dagger }\right\rangle \right\rangle _\omega $denotes the orthogonal
Green function in Bloch representation. The Fourier transformation of the
Green function $G_{ij}(\omega )$ is then seen to be

\begin{equation}
G_{ij}(\omega )=\frac 1{N_s}\sum_ke^{i{\bf k\cdot (x}_i-{\bf x}%
_j)}G_k(\omega )
\end{equation}
Substituting Eq.(22) into Eq.(20) the single particle Green function in the
Bloch representation is explicitly obtained as

\begin{equation}
G_k(\omega )=\frac{\omega -\varepsilon +Un_0}{(\omega -\varepsilon
(k))(\omega -\varepsilon -Un_0)-Un_0(\varepsilon (k)-\varepsilon )}
\label{result1}
\end{equation}
Rewrite the solution as the following form

\begin{equation}
G_k(\omega )=\frac{A_k^{(1)}}{\omega -E^{(1)}}+\frac{A_k^{(2)}}{\omega
-E_k^{(2)}}  \label{result2}
\end{equation}
where $E^{(1)}$ and $E_k^{(2)}$ denote the poles of the Green function $%
G_k(\omega )$, and it is seen that the excitation spectrum possesses a band
structure such as

\begin{equation}
E^{(1)}=\varepsilon \text{, }E_k^{(2)}=\varepsilon (k)+Un_0
\label{last result}
\end{equation}
The lowest band shrinks to a single level of zero band width (see Fig.1).
Although the energy spectrum Eq.(\ref{last result}) comprises two parts,
they may, in a certain case depending on the relative values of the
interatomic repulsion $U$ and the tunnel coupling $J$, be merged into the
same energy band. The energy gap between the two bands is [Fig.1]

\[
\Delta =E_k^{(2)}\left| _{k=0}\right. -E^{(1)}=Un_0-Jz 
\]

When the constant of the interatomic repulsion $U$ is large with respect to
the tunnel coupling $J$ such that $\Delta >0$, a gap exists implying the
MIP. With increasing of tunnel coupling $J$ the gap width $\Delta $
decreases and finally the two energy bands in the excitation spectrum
overlap and the gap disappears, indicating the SFP. We then obtain the
condition of SMI phase transition that

\begin{equation}
\Delta =0
\end{equation}
,namely, 
\begin{equation}
\frac U{zJ}=\frac 1{n_0}  \label{condition}
\end{equation}
which agrees with the result in refs.\cite{D. van Oosten}\cite{Matthew P. A.
Fisher} \cite{Freericks}\cite{Sheshadri}.

To see the SFP more closely we take the zero wave vector limit of the energy
band $E_k^{(2)}$ ( $k\rightarrow 0)$

\[
E_k^{(2)}\sim \varepsilon -Jz+Un_0+\frac 1{2^3}Jz\lambda ^2k^2. 
\]
Under the condition Eq.(\ref{condition}) at which the energy gap between $%
E^{(1)}$ and $E_k^{(2)}$ disappears, we have a gapless Goldstone mode in the
excitation spectrum such as

\begin{equation}
E_{exc}\sim \frac 1{2^3}Jz\lambda ^2k^2
\end{equation}
which is different from the result of Bogoliubov theory for the system of
weakly interacting bosons, in the absence of the periodic potential, where
the wave vector dependence of the excitation spectrum is linear in the zero
wave vector limit so that a non-vanishing velocity can exist. Strictly
speaking what we obtained here is an ordinary fluid phase. The energy
spectrum of Eq.(25) determined with the help of Green function method is too
simple to realize the superfluid phase explicitly. This may be due to the
particular procedure of the approximation used in the above derivation. It
is certainly of interest to study the spectrum of bosonic atoms in the BEC
trapped in the optical lattice in terms of Bogliubov method to see weather
or not the system can possess a superfluid phase which we are going to
discuss in the next section.

It is worth while to point out that when the interaction between bosons
vanishes $i$.e. $U=0$, the Green function Eq.(\ref{result1}) reduces to the
well known single band solution

\begin{equation}
G_k(\omega )\left| _{U=0}\right. =\frac 1{\omega -\varepsilon \left(
k\right) }
\end{equation}
for bosons in a periodic potential.

\section{BOGLIUBOV METHOD}

Now we study the energy spectrum of boson atoms in the optical lattice by
means of the Bogliubov method . Using the relation(\ref{transform}) the
Hamiltonian(\ref{Hamiltonian1}) can be converted to 
\[
\widehat{H}=\sum_k\varepsilon (k)\widehat{a}_k^{\dagger }\widehat{a}_k+%
\widehat{H}_{int} 
\]

\begin{equation}
\widehat{H}_{int}=\frac U{2N_s}\sum_{k,p,k^{\prime },p^{\prime }}\delta
_{k+p,k^{\prime }+p^{\prime }}\widehat{a}_{k^{\prime }}^{\dagger }\widehat{a}%
_{p^{\prime }}^{\dagger }\widehat{a}_k\widehat{a}_p\quad \text{.}
\label{interation2}
\end{equation}
Since the number of atoms condensed in the zero-momentum state is much
larger than one, we have $\widehat{a}_0\widehat{a}_0^{\dagger }=\widehat{a}%
_0^{\dagger }\widehat{a}_0+1\simeq N_0\gg 1$, where $N_0$ is the total
number of condensed atoms. Thus we can replace the operator $\widehat{a}_0$
and $\widehat{a}_0^{\dagger }$ by a ''$c"$ number $\sqrt{N_0}$. The
interacting part of the Hamiltonian (\ref{interation2}) can be written as
(in the order of $N_0$)

\[
\widehat{H}_{int}=\frac U{2N_s}N_0^2+\frac{UN_0}{2N_s}\sum\nolimits_k^{^{%
\prime }}(\widehat{a}_k\widehat{a}_{-k}+\widehat{a}_k^{\dagger }\widehat{a}%
_{-k}^{\dagger }+2\widehat{a}_k^{\dagger }\widehat{a}_k) 
\]
and the total Hamiltonian is

\begin{equation}
\widehat{H}=\frac{UN_0^2}{2N_s}+N_0\left( \varepsilon -zJ\right)
+\sum\nolimits_k^{^{\prime }}\left\{ \frac{Un_0}2(\widehat{a}_k\widehat{a}%
_{-k}+\widehat{a}_k^{\dagger }\widehat{a}_{-k}^{\dagger })+\left(
\varepsilon (k)+Un_0\right) \widehat{a}_k^{\dagger }\widehat{a}_k\right\}
\label{Hamiltonian3}
\end{equation}
where $\sum\nolimits_k^{^{\prime }}$denotes the sum with exclusion of the
term of $k=0$.

The following Bogoliubov transformation is introduced in order to
diagonalize the Hamiltonian(\ref{Hamiltonian3})

\begin{equation}
\left\{ 
\begin{array}{c}
\widehat{b}_k=u_k\widehat{a}_k+v_k\widehat{a}_{-k}^{\dagger } \\ 
\widehat{b}_k^{\dagger }=u_k\widehat{a}_k^{\dagger }+v_k\widehat{a}%
_{-k}\quad \text{.}
\end{array}
\right. \quad
\end{equation}
We require that the quasi-boson operators $\widehat{b}_k$ and $\widehat{b}%
_k^{\dagger }$ satisfy the usual commutation relation, $[\widehat{b}_k$, $%
\widehat{b}_k^{\dagger }]=1$ , which leads to the condition

\begin{equation}
u_k^2-v_k^2=1  \label{nomalization}
\end{equation}
for the coefficients $u_k$ and $v_k$ and then the Hamiltonian can be written
as

\[
\widehat{H}=E_c+\widehat{H}_1+\widehat{H}_2 
\]
where

\begin{equation}
E_c=\frac 12UN_0n_0+N_0(\varepsilon -zJ)  \label{E_0}
\end{equation}
is a constant and

\begin{equation}
\widehat{H}_1=\sum\nolimits_k^{^{\prime }}\left( \left( u_k^2+v_k^2\right)
\left( \overline{\varepsilon }_k+Un_0\right) -2Un_0u_kv_k\right) \widehat{b}%
_k^{\dagger }\widehat{b}_k
\end{equation}

\begin{equation}
\widehat{H}_2=\sum\nolimits_k^{^{\prime }}\left( \frac{Un_0}2\left(
u_k^2+v_k^2\right) -\left( \overline{\varepsilon }_k+Un_0\right)
u_kv_k\right) \left( \widehat{b}_k\widehat{b}_{-k}+\widehat{b}_k^{\dagger }%
\widehat{b}_{-k}^{\dagger }\right)
\end{equation}
To eliminate the off-diagonal part $\widehat{H}_2$ we require

\begin{equation}
\frac{Un_0}2\left( u_k^2+v_k^2\right) -\left( \overline{\varepsilon }%
_k+Un_0\right) u_kv_k=0  \label{calcellation}
\end{equation}
where $\overline{\varepsilon }_k=zJ(1-\cos \left( kd\right) ).$ Introducing
a parameter $\phi _k$ such that

\[
u_k=\cosh \phi _k\qquad v_k=\sinh \phi _k 
\]
the conditions (\ref{nomalization}), (\ref{calcellation}) lead to the useful
relations

\[
\tanh 2\phi _k=\frac{2u_kv_k}{u_k^2+v_k^2}=\frac{Un_0}{\overline{\varepsilon 
}_k+Un_0}
\]

\[
u_k^2+v_k^2=\cosh (2\phi _k)=\frac{\overline{\varepsilon }_k+Un_0}{E_k} 
\]
with which the diagonalized Hamiltonian is obtained as 
\[
\widehat{H}=E_c+\sum\nolimits_k^{^{\prime }}E_k\widehat{b}_k^{\dagger }%
\widehat{b}_k 
\]
where the energy spectrum $E_k$ of quasi-particle is

\begin{equation}
E_k=\sqrt{\overline{\varepsilon }_k\left( \overline{\varepsilon }%
_k+2Un_0\right) }\text{.}  \label{energy specturm2}
\end{equation}
The energy spectrum is different from that of Eq.(25) and is typical for the
superfluid. The energy gap $\Delta _g$ of excitation spectrum is obviously

\begin{eqnarray}
\Delta _g &=&\frac{E_c-N_0\varepsilon }{N_0}  \nonumber \\
&=&\frac 12Un_0-zJ\text{ .}  \label{delta}
\end{eqnarray}
The phase transition condition determined from $\Delta _g=0$ is

\begin{equation}
\frac U{2zJ}=\frac 1{n_0}\text{ }
\end{equation}
which shows a factor 2 difference comparing with the condition in Eq.(\ref
{condition}). This may be caused by the approximation itself. When the
energy gap disappears, i.e. $\Delta _g=0$, the dispersion relation of $E_k$($%
k\rightarrow 0)$ reads

\begin{equation}
E_k\backsim \left( zJUn_0d^2\right) ^{1/2}k
\end{equation}
indicating explicitly the superfluididy in agreement with the Bogliubov
superfluid theory for weakly interacting bosons in the absence of the
periodic potential. The linear wave vector dependence of the excitation
spectrum $E_k$ ( unlike the ordinary fluid Eq.(28) where $E_{exc}$ is
proportional to $k^2$) is the characteristic of the sperfluid which gives
rise to a persistent velocity of superfluid or quasi-particle found as 
\begin{equation}
v_s=\left( \frac{\partial E_k}{\partial k}\right) _{k\rightarrow 0}=\left(
zJUn_0d^2\right) ^{1/2}\text{.}  \label{velocity}
\end{equation}
\quad For the case of boson atoms with repulsive interaction($a_s>0$), the
parameters $J$ and $U$ are positive, $v_s$ is a real number which implies a
persistent current. The velocity $v_s$ can be controlled by the tuning of
laser lights which result in the optical lattice. As seen from the
definitions (\ref{J}), (\ref{U}) for $J$ and $U$, these parameters both
depend on the Wannier functions which are essentially determined by the
potential of optical lattice. Therefore $J$ and $U$ are not independently
tunable by the adjusting of the laser parameters. In fact when the depth of
the lattice potential increases, the hopping matrix element $J$ decreases
exponentially while the matrix- element of the on-site interaction, $U$ ,
increases. We thereby expect that there exists a maximum value of the
persistent velocity $v_s$ in some particular values of $J$ and $U$.

\section{CONCLUSION}

We have studied the Bose-Hubbard model of BEC trapped in a periodic
potential in terms of Green function method and Bogliubov transformation as
well. The condition of phase transition between SFP and MIP is determined by
the energy band structure of the excitation spectrum due to, obviously, the
competition between the interatomic repulsion and the tunnel coupling. Our
result agrees with the condition of SMI phase transition obtained in
literature. The SFP property of BEC in the optical lattice is explained
explicitly from energy spectrum derived by means of the Bogliubov approach.
It is shown that the persistent velocity of the quasi-particle in SFP can be
tuned by the adjusting of the laser lights which result in the optical
lattice.

This work was supported by the Youth Science Foundations of Shanxi Province
and Shanxi University, and by the National Science Foundation of China under
Grants Nos. 10075032 and 10174095.

\begin{description}
\item  {\large {Figure captions}}

\item[Fig.~1.]  Excitation spectrum and energy gap.
\end{description}

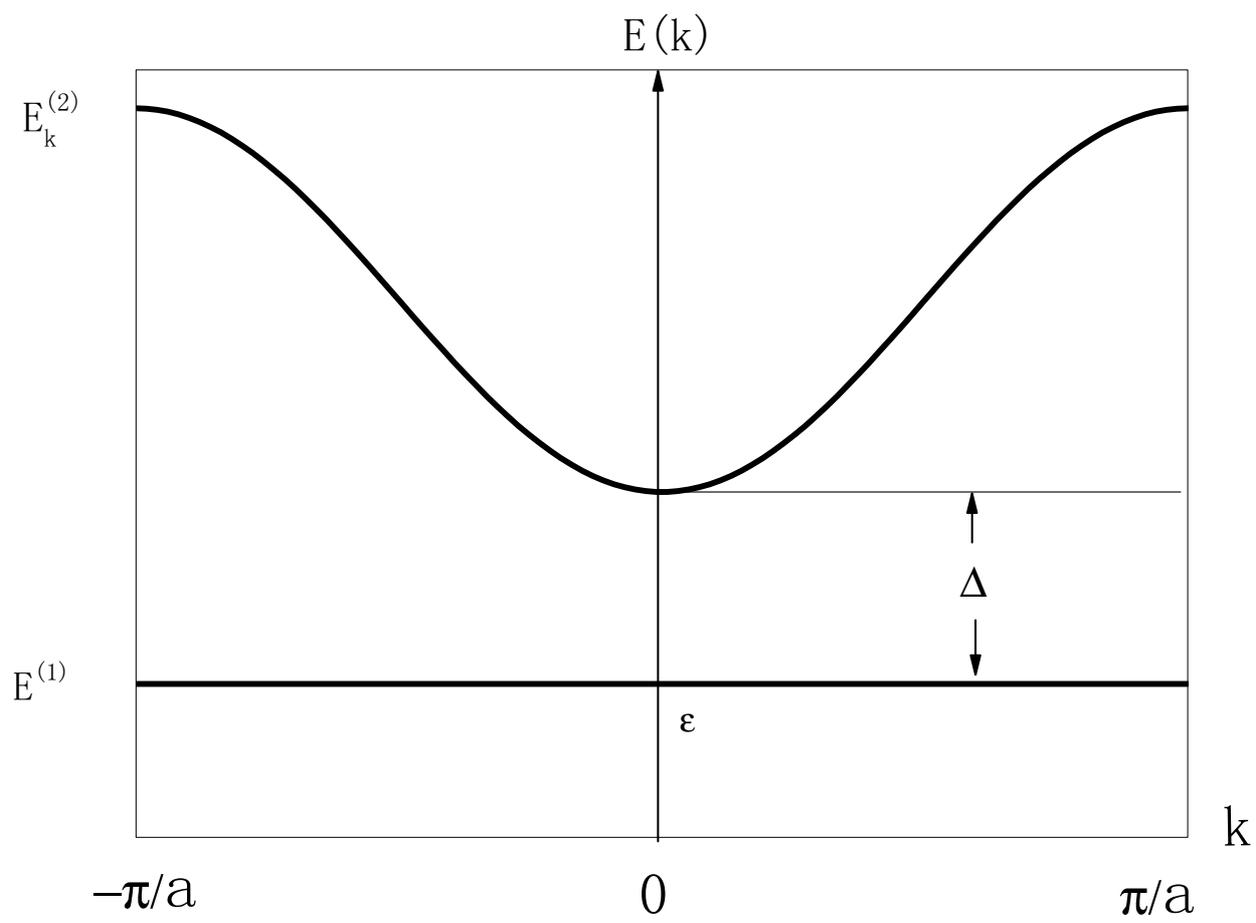
\end{document}